# Deterministic coupling of delta-doped NV centers to a nanobeam photonic crystal cavity

Jonathan C. Lee[1)], David O. Bracher[2)], Shanying Cui[1)], Kenichi Ohno[3)], Claire A. McLellan[3)], Xingyu Zhang[1)], Paolo Andrich[3,4)], Benjamin Alemán[3,a)], Kasey J. Russell[1)], Andrew P. Magyar[1)], Igor Aharonovich[1,5)], Ania Bleszynski Jayich[3)], David Awschalom[3,4)], Evelyn L. Hu[1,b)]

[1]School of Engineering and Applied Sciences, Harvard University, Cambridge, MA 02138, USA

[2]Department of Physics, Harvard University, Cambridge, MA 02138, USA

[3]Department of Physics, University of California, Santa Barbara, Santa Barbara, CA 93106, USA.

[4]Institute for Molecular Engineering, University of Chicago, Chicago, IL 60637, USA

[5]School of Physics and Advanced Materials, University of Technology Sydney, Ultimo, NSW 2007, Australia

The negatively-charged nitrogen vacancy center (NV) in diamond has generated significant interest as a platform for quantum information processing and sensing in the solid state. For most applications, high quality optical cavities are required to enhance the NV zero-phonon line (ZPL) emission. An outstanding challenge in maximizing the degree of NV-cavity coupling is the deterministic placement of NVs within the cavity. Here, we report photonic crystal nanobeam cavities coupled to NVs incorporated by a delta-doping technique that allows nanometer-scale vertical positioning of the emitters. We demonstrate cavities with Q up to ~24,000 and mode volume $V \sim 0.47(\lambda/n)^3$ as well as resonant enhancement of the ZPL of an NV ensemble with Purcell factor of ~20. Our fabrication technique provides a first step towards deterministic NV-cavity coupling using spatial control of the emitters.

A diamond-based emitter-cavity system provides an important platform for the realization of quantum information processing and sensing in the solid state[1–4]. The long electron spin coherence of the negatively-charged nitrogen vacancy center (subsequently referred to as NV) in diamond together with the ability to control and read out the spin optically make the NV an exceptional emitter and qubit[5,6]. Recent advances in diamond nanofabrication have enabled fabrication of high quality two dimensional photonic crystal cavities and microring resonators that can couple to and enhance NV emission[7–13]. Nevertheless, a major remaining technical challenge for emitter-cavity coupling is the deterministic placement of the NVs within the cavity. This article reports the incorporation of NVs formed by a delta-doping growth technique[14] that allows nanometer-scale vertical positioning of the emitters within the cavity. This is important to achieve maximum field enhancement between the emitter and the cavity mode. The delta-doped NVs were integrated into photonic crystal nanobeam cavities, with quality factor Q as high as ~ 24,000 and a small mode volume of V ~ 0.47 $(\lambda/n)^3$. Resonant enhancement of the zero-phonon line (ZPL) of an ensemble of NVs is observed with Purcell factor of ~ 20. The fabrication method introduced here provides a first step toward using spatial control to enable deterministic coupling of NV centers.

Resonant enhancement of the NV ZPL transition can be achieved within high-quality optical cavities, where the enhancement is proportional to the well-known Purcell factor. When the NV ZPL and the cavity are resonant in frequency, and the NV is ideally positioned with respect to the cavity field (mode), we can express the Purcell factor[2,15] as

$$F \approx \frac{3}{4\pi^2}\left[\frac{\lambda}{n}\right]^3 \frac{Q}{V}, \qquad [1]$$

where λ is the resonant wavelength and n the index of refraction of the cavity material.

While the cavity can be designed to achieve spectral overlap with the NV, it is more challenging to realize the spatial overlap between the NV and the field maximum of the cavity mode necessary for ideal coupling. The current work addresses part of this challenge by demonstrating high Q/V cavities with integrated NV emitters located at a controlled vertical position within the cavity.

The cavity design used in these experiments is a photonic crystal nanobeam with a linearly tapered lattice constant in the middle of the cavity (dotted line region in Figure 1.a), sandwiched by a Bragg mirror on each side.[16] This design was chosen to achieve a high Q and small mode volume. The tapered lattice constant leads to the localization of the cavity mode, as described in previous references.[17] The electric field intensity profile for the fundamental transverse electric (TE) mode is shown in Figures 1.d and e, and was simulated using finite-difference time-domain (FDTD) simulation software (Lumerical Solutions, Inc.). The theoretical Q for the fundamental TE mode of the cavity is ~ 270,000 and the mode volume is ~ 0.47 $(\lambda/n)^3$. A representative spectrum showing this mode decorating NV center luminescence is shown in Figure 1.f.

The fabrication of the cavities, outlined in Figure 2, starts with a CVD-grown (100) single-crystal diamond substrate from Element Six ™.[16] An ion-damaged layer is generated using a high energy $He^+$ implantation (1 MeV, $5\times10^{16}$ $cm^{-2}$). The damaged layer allows for removal and lift-off of a ~ 1.7 μm-thick diamond membrane[18]. The delta-doped layer is then overgrown on the diamond substrate using plasma enhanced chemical vapor deposition (PECVD, Seki Technotron), with growth conditions as previously described.[14] The final overgrown diamond film comprises two 100 nm buffer layers sandwiching a ~ 6 nm-thick nitrogen delta-doped layer. An electrochemical etch process selectively attacks the material at the peak of the ion damage, lifting-off the membrane from the diamond substrate. The diamond membranes are then stamped onto a poly-methyl methacrylate (PMMA)-coated silicon sample, with the delta-doped film in contact with the PMMA. To remove the damaged material, the diamond membrane is thinned, using oxygen-based, inductively-coupled plasma reactive ion etching (ICP-RIE, Unaxis

Shuttleline). This produces a delta-doped membrane ~ 200 nm thick. The linewidth of the diamond Raman peak measured from the membrane (FWHM 2.14 ± 0.05 cm$^{-1}$) is comparable to that of the starting bulk diamond (FWHM 2.43 ± 0.10 cm$^{-1}$).

Finally, the cavities are fabricated from the delta-doped membranes using electron beam lithography and a negative-tone hydrogen silsesquioxane (HSQ) based resist (Dow-Corning XR-1541). After development, the HSQ-based resist serves as an etch mask for the subsequent oxygen-based ICP-RIE diamond etch step (see Fig 2.h), which also undercuts the structures by selectively removing the PMMA bonding layer beneath the cavities.

SEM micrographs of the diamond nanobeam cavity are shown in Figure 1.a-c. The experimental cavity Q can be estimated by fitting the cavity mode from the measured photoluminescence (PL) spectrum using a Lorentzian curve ($Q = \lambda_{cav}/\Delta\lambda_{cav}$ where $\Delta\lambda_{cav}$ is the FWHM of the Lorentzian). The highest Q measured from the photonic crystal cavities is ~ 24,000. This is lower than simulated Q values due to non-idealities in fabrication. Nevertheless, these devices are distinctive in their high quality factors and engineered low mode volumes.

To demonstrate resonant enhancement of the NV centers within the photonic crystal nanobeam, the frequency of the cavity mode is tuned to match that of the NV emission. The cavity mode is red-shifted in steps of ~ 0.07 nm by injecting increments of nitrogen gas into the cryostat where the cavity was maintained at a temperature of 4.5 K.[16] As the cavity mode is tuned into resonance with the NV ZPL, we observe both an increase in fluorescence intensity and a decrease in fluorescence lifetime. Figure 3 shows the systematic tuning of a mode, with Q of 7,000 and starting wavelength 636 nm, into and then out of resonance with the NV ZPL (637.8 nm). The fluorescence signal from the NV phonon sideband is filtered out using a band-pass filter. When the cavity mode is tuned into resonance with the NV ZPL (step 54 in Figure 3.b), the PL intensity increased by a factor of 27 compared to the off-resonance PL intensity (step 28 in Figure 3.b), indicating a Purcell factor of 26 (ref. 10).

The Purcell enhancement factor may also be deduced from lifetime measurements. These measurements are carried out using an avalanche photodiode (APD, Micro Photon Device) to produce spectra of fluorescence decay which are fit with a double exponential decay model (see Fig 3.c). The NV lifetime measured from a diamond membrane, without any cavity structure, is $12.96 \pm .7$ ns, with a short decay component of $0.22 \pm 0.03$ ns. The short decay component is attributed to the background fluorescence from other defects in the diamond membrane. As the cavity mode spectrally approaches the NV ZPL, the lifetime is shortened, with a long decay component of $\tau_{on-resonance} = 10.43 \pm 0.5$ ns, with short decay component of $2.18 \pm .4$ ns. Additionally, for the cavity structure, the off-resonance NV lifetime is $\tau_{off-resonance} = 22.34 \pm 1.1$ ns, and the short decay component is $0.19 \pm 0.03$ ns. This longer lifetime is consistent with the lowered density of optical states for energies within the photonic band gap[19]. Following the method of Faraon et al.[10], we can use the on- and off-resonance lifetimes to estimate the Purcell enhancement factor. The observed NV lifetime is due to the contributions of both the ZPL and the phonon sideband: $\frac{1}{\tau_{NV}} = \frac{1}{\tau_{ZPL}} + \frac{1}{\tau_{PS}}$. In the off-resonance scenario, the decay rate of NV ($\frac{1}{\tau_{NV}}$) is approximately equal to the decay rate to sideband ($\frac{1}{\tau_{PS}}$), due to the small branching ratio into the ZPL. On resonance, the ZPL decay rate ($\frac{1}{\tau_{ZPL}}$) undergoes Purcell enhancement, and it is assumed that the decay rate from sideband is the same as was observed while off resonance. The two cases can then be combined to yield a value for the factor of Purcell enhancement, F

$$F = \tau_{ZPL,0} \left( \frac{1}{\tau_{on-resonance}} - \frac{1}{\tau_{off-resonance}} \right). \quad [2]$$

Here, $\tau_{ZPL,0}$ is the ZPL lifetime in the diamond membrane, given by the ratio of the membrane NV lifetime to the branching ratio. Assuming a branching ratio of 0.03,[10] we find a Purcell enhancement factor of 22.

The enhanced intensity and lifetime reduction are manifestations of the Purcell effect and indicate good optical coupling of NV centers with high-Q optical cavities. However, while our cavities have the highest Q/V ratio yet reported for diamond, our Purcell enhancement factors are not correspondingly higher. Our intensity increase is ~4 times greater than that reported in Hausmann et al., despite a Q/V ratio ~30 times greater[11]. Similarly Faraon et. al report a Purcell factor of 70, as calculated from lifetime reduction, whereas we find a factor of just 22, despite a ~5 times greater Q/V.[10] One explanation for this discrepancy may be the presence of several NV centers in close proximity to the cavity mode. The best fit of the NV ZPL in the nanobeam suggests an ensemble of several NVs with FWHMs in the range of 0.2-0.35 nm. The cavity mode is much narrower than the linewidths of the NVs, and the mode is therefore not likely fully coupled to the NV ZPL. Furthermore, if the NV is not perfectly positioned and aligned with the cavity field antinode, it is not possible to achieve the degree of coupling suggested by the Q/V ratio alone[2,12].

We believe that the controlled placement of NVs in the vertical direction within a high-quality diamond cavity marks an important step in maximizing NV-cavity coupling. In addition, three-dimensional positioning of NVs[20] within a cavity and control of the NV alignment[21–23] would allow further improvements in Purcell enhancement and device yield.

In this work we have demonstrated resonant enhancement of the ZPL from NV centers located at pre-determined vertical positions coupled to a photonic crystal cavity with Q factor ~ 7,000 and mode volume ~ 0.47 $(\lambda/n)^3$. Using a $N_2$ gas condensation method in a cryogenic environment, we were able to tune the cavity mode through a range of ~ 10 nm to match the frequency of the cavity resonance to that of the ZPL emission. Greater than 20-fold Purcell enhancement of the NV ZPL is inferred from the lifetime modification observed from NV centers while on and off-resonance with the cavity mode. Nanobeam photonic crystal cavities with Q-factors as high as 24,000 were also fabricated using these diamond membranes with delta-doped NV layers. This work provides a method to deterministically couple NV centers to high quality photonic crystal cavities and helps pave the way for scalable quantum information processing using NV centers in diamond.

**Acknowledgements:**
The authors thank D. R. Clarke for access to confocal Raman microscope, A. Woolf for help with cavity tuning, M. Huang for assistance with ion implantation, and T.L. Liu and Y. Zhang for helpful discussions. I. A. is the recipient of an Australian Research Council Discovery Early Career Research Award (Project No. DE130100592). C.A.M. and S.C. are supported by an NSF Graduate Research Fellowship. The authors acknowledge funding from the Air Force Office of Scientific Research, under the QUMPASS program (Award No. FA9550-12-1-0004) and the STC Center for Integrated Quantum Materials, NSF grant DMR-1231319. This work was performed in part at the Center for Nanoscale Systems (CNS), a member of the National Nanotechnology Infrastructure Network (NNIN), which is supported by the National Science Foundation under NSF award no. ECS-0335765. CNS is part of Harvard University.

**Materials and Methods**

*Cavity design*

Finite-difference time-domain (FDTD) simulations were carried out using commercially available software (FDTD Solutions, Lumerical Solutions, Inc.) to model the photonic crystal cavities. The nanobeam cavities had lattice constant $a = 213$ nm, cavity thickness $h = 200$ nm, and hole radius $r = 0.32*a$. The linearly tapered region consists of 4 lattice points on each side where the innermost lattice constant is reduced to 0.84*a. The mesh size used in this simulation is a/32 to accurately estimate the mode volume, and the simulation time is 12 ps for the high-quality factor design.

*Fabrication Process*

The initial substrate is an electronic grade CVD diamond with (100) orientation from Element Six[TM]. An ion-damaged layer is generated using a high energy He$^+$ implantation (1 MeV, $5\times10^{16}$ cm$^{-2}$). The delta-doped nitrogen layer is then grown using plasma-enhanced chemical vapor deposition (PECVD, AsTex) designed to create a 100 nm-thick buffer layer, a ~ 6 nm-thick delta-doped layer, and a 100 nm-thick cap layer. We used naturally abundant $N_2$ gas for doping with nitrogen. The growth conditions are: $CH_4/H_2$ = 0.025% at a pressure of 25 Torr with 750 W RF power using isotopically purified[12] $CH_4$. To form NV centers in the delta-doped layer, high-energy electron irradiation (2 MeV, $5\times10^{14}$ cm$^{-2}$) is performed on the sample to create vacancies, which is followed by thermal annealing (850 °C, in $H_2$/Ar forming gas for 2 hours). The diamond membranes with delta-doped NV centers are lifted off using an electrochemical etch to remove the ion-damaged layer by immersing the sample in ultra-pure water with a 12 V bias.

*Optical characterization*

We used two different confocal microscopes to measure the photoluminescence (PL) signal in this paper. One is a commercially available Raman confocal microscope (Horiba Jobin-Yvonne). The confocal microscope consists of a high numerical aperture objective (NA = 0.9) with 100× magnification mounted perpendicular to the sample. The excitation source is a 532 nm diode laser. The PL signal is collected using the same objective that focuses the excitation laser through a pinhole (50 μm) at the focal point in the back of the objective. The collected light is sent to a spectrometer with a charge-coupled device (CCD) camera at the end of the spectrometer.

The other is a home-built confocal microscope with a cryostat that is used for the gas-tuning experiment and lifetime measurements. This confocal setup consists of a coverslip-corrected objective with N.A = 0.6 and 40× magnification (Olympus) mounted perpendicular to the sample and is used for both excitation and collection. The excitation source used for this setup is a super continuum laser with pulse rate of 76 MHz, filtered to select only 530-536 nm for excitation. A multi-mode fiber with core diameter of 25 μm used for signal collection is connected to a spectrometer with a liquid nitrogen-cooled CCD camera or avalanche photodiode (APD). The NV lifetime is fit using a geometric series of exponential decay where the signal is deconvolved with the instrument response function (IRF) of the avalanche photodiode. During the cavity tuning process, the bandpass filter used has a center wavelength of 636 nm with FWHM bandwidth of 14.6 nm and is manufactured by Semrock.

*Gas tuning*

Nitrogen gas is stored in a small chamber at a controlled pressure before it is injected into the cryostat through a gas valve and a metal tube. The gas flow into the cryostat is controlled by the pressure in the tube before its injection into the cryostat: the higher the pressure, the larger the flow. By using a 1 Torr pressure, we can tune the cavity mode by ~ 0.07 nm/step, and we can tune 0.5 nm/step using 10 Torr pressure.

**References:**

[1]D. D. Awschalom, L. C. Bassett, A. S. Dzurak, E. L. Hu, and J. R. Petta, Science 339, 1174 (2013).
[2]I. Aharonovich, A. D. Greentree, and S. Prawer, Nat. Photonics 5, 397 (2011).
[3]M. Lončar and A. Faraon, MRS Bull. 38, 144 (2013).

[4]W. Pfaff, B. Hensen, H. Bernien, S. B. van Dam, M. S. Blok, T. H. Taminiau, M. J. Tiggelman, R. N. Schouten, M. Markham, D. J. Twitchen, R. Hanson, Science 345, 532 (2014).
[5]P. C. Maurer, G. Kucsko, C. Latta, L. Jiang, N. Y. Yao, S. D. Bennett, F. Pastawski, D. Hunger, N. Chisholm, M. Markham, D. J. Twitchen, J. I. Cirac, and M. D. Lukin, Science 336, 1283 (2012).
[6]H. Bernien, B. Hensen, W. Pfaff, G. Koolstra, M. S. Blok, L. Robledo, T. H. Taminiau, M. Markham, D. J. Twitchen, L. Childress, and R. Hanson, Nature 497, 86 (2013).
[7]I. Aharonovich, J. C. Lee, A. P. Magyar, B. B. Buckley, C. G. Yale, D. D. Awschalom, and E. L. Hu, Adv. Mat. 24, OP54 (2012).
[8]A. Faraon, P. E. Barclay, C. Santori, K.-M. C. Fu, and R. G. Beausoleil, Nat. Photonics 5, 301 (2011).
[9]B. J. M. Hausmann, B. Shields, Q. Quan, P. Maletinsky, M. McCutcheon, J. T. Choy, T. M. Babinec, A. Kubanek, A. Yacoby, M. D. Lukin, and M. Lončar, Nano Lett. 12, 1578 (2012).
[10]A. Faraon, C. Santori, Z. Huang, V. M. Acosta, and R. G. Beausoleil, Phys. Rev. Lett. 109, 033604 (2012).
[11]B. J. M. Hausmann, B. J. Shields, Q. Quan, Y. Chu, N. P. de Leon, R. Evans, M. J. Burek, A. S. Zibrov, M. Markham, D. J. Twitchen, H. Park, M. D. Lukin, and M. Lončar, Nano Lett. 13, 5791 (2013).
[12]D. Englund, B. Shields, K. Rivoire, F. Hatami, J. Vuckovic, H. Park, and M. D. Lukin, Nano Lett. 10, 3922 (2010).
[13]J. Riedrich-Möller, C. Arend, C. Pauly, F. Mücklich, M. Fischer, S. Gsell, M. Schreck, and C. Becher, Nano Lett. 14, 5281 (2014).
[14]K. Ohno, F. J. Heremans, L. C. Bassett, B. A. Myers, D. M. Toyli, A. C. Bleszynski Jayich, C. J. Palmstrøm, and D. D. Awschalom, Appl. Phys. Lett. 101, 082413 (2012).
[15]E. M. Purcell, Phys. Rev. 69, 681 (1946).
[16]See supplemental materials for more information on cavity design, diamond fabrication, optical characterization, and gas tuning
[17]Y. Zhang, M. W. McCutcheon, I. B. Burgess, and M. Lončar, Opt. Lett. 34, 2694 (2009).
[18]A. P. Magyar, J. C. Lee, A. M. Limarga, I. Aharonovich, F. Rol, D. R. Clarke, M. Huang, and E. L. Hu, Appl. Phys. Lett. 99, 081913 (2011).
[19]E. Yablonovitch, Phys. Rev. Lett. 58, 2059 (1987).
[20]K. Ohno, F. J. Heremans, C. F. de las Casas, B. A. Myers, B. J. Alemán, A. C. Bleszynski Jayich, and D. D. Awschalom Appl. Phys. Lett. 105, 052406 (2014).
[21]J. Michl, T. Teraji, S. Zaiser, I. Jakobi, G. Waldherr, F. Dolde, P. Neumann, M. W. Doherty, N. B. Manson, J. Isoya, and J. Wrachtrup, Appl. Phys. Lett. 104, 102407 (2014).
[22]M. Lesik, J.-P. Tetienne, A. Tallaire, J. Achard, V. Mille, A. Gicquel, J.-F. Roch, and V. Jacques, Appl. Phys. Lett. 104, 113107 (2014).
[23]E. Neu, P. Appel, M. Ganzhorn, J. Miguel-Sánchez, M. Lesik, V. Mille, V. Jacques, A. Tallaire, J. Achard, and P. Maletinsky, Appl. Phys. Lett. 104, 153108 (2014).

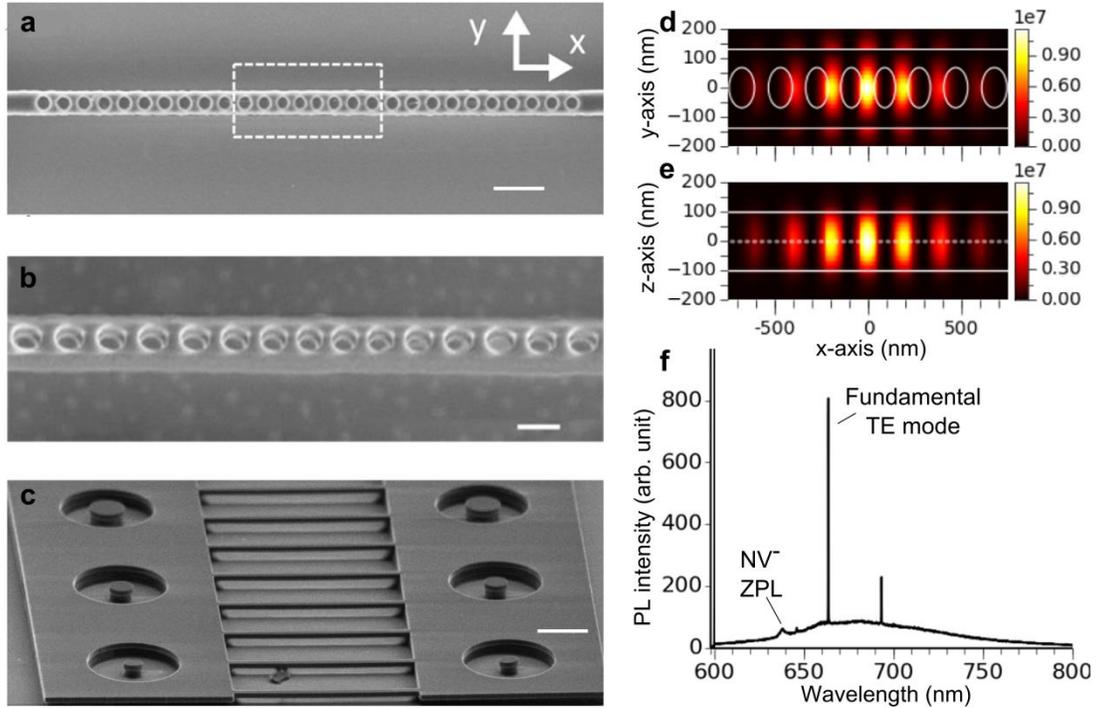

**Figure 1. Diamond photonic crystal cavities**. **a.** SEM viewgraph of the diamond nanobeam cavity. The dashed box highlights the defect region, for which a simulation is shown in (d) and (e). Scale bar indicates 500 nm. **b.** A magnified view of the nanobeam, showing the defect region more closely. Scale bar indicates 200 nm. **c.** An array of nanobeams, showing beam undercut. Scale bar indicates 2 μm. **d.** Finite-difference time-domain (FDTD) simulation of defect region highlighted in (a) showing the electric field intensity profile in the x-y plane for the fundamental transverse electric (TE) mode. The white line represents the edges of the nanobeam cavity. **e.** FDTD simulation of the defect region highlighted in (a) showing the electric field intensity profile for the fundamental TE mode in the x-z plane. The dotted white line denotes the designed delta-doped NV layer, showing that the delta-doped layer is near the field maximum in the z-direction. **f.** A representative photoluminescence (PL) measurement showing a representative nanobeam spectrum. NV centers within the beam are excited by a 532 nm diode laser, as seen by the NV$^-$ zero phonon line (ZPL) along with the phonon sideband. The NV luminescence is decorated by the nanobeam modes, including the fundamental TE mode as labeled.

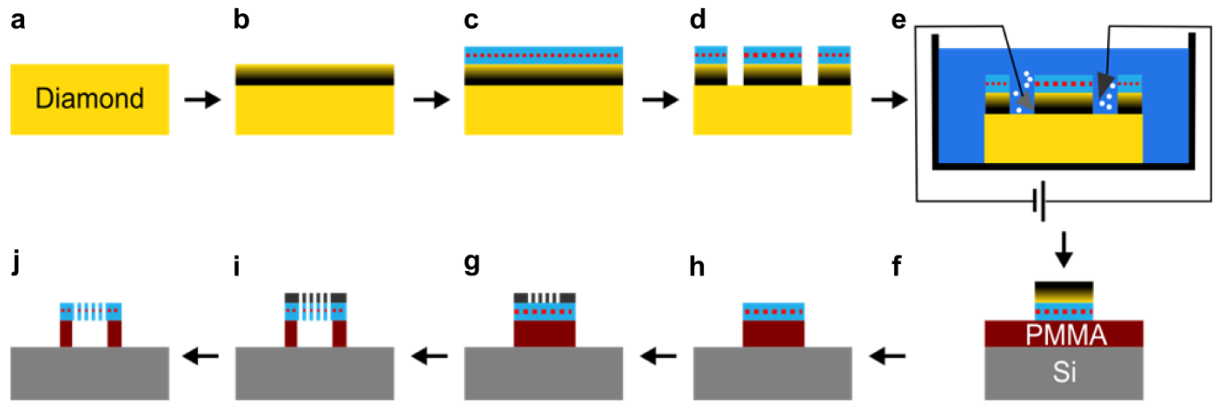

**Figure 2. Fabrication of diamond photonic crystal cavities with delta-doped NV layer. a.** The starting material is a CVD diamond. **b.** A helium ion implantation is performed on the bulk diamond sample, followed by thermal annealing to create an ion-damaged layer. **c.** A diamond thin film is grown on the implanted diamond with a delta-doped NV layer. **d.** Mesa structures are patterned on bulk diamond using photolithography and reactive ion etching. **e.** The membranes are lifted off in an aqueous solution with an applied bias. **f.** The membranes are then stamped on PMMA-coated Si samples. **g.** The unwanted ion-damaged layer is removed using ICP-RIE. **h.** Nanobeam photonic crystal cavities are patterned using e-beam lithography. **i.** $O_2$ ICP-RIE is used to etch the diamond with a slight over-etching to undercut the PMMA adhesion layer. **j.** The HSQ-based mask is removed using buffered oxide etchant.

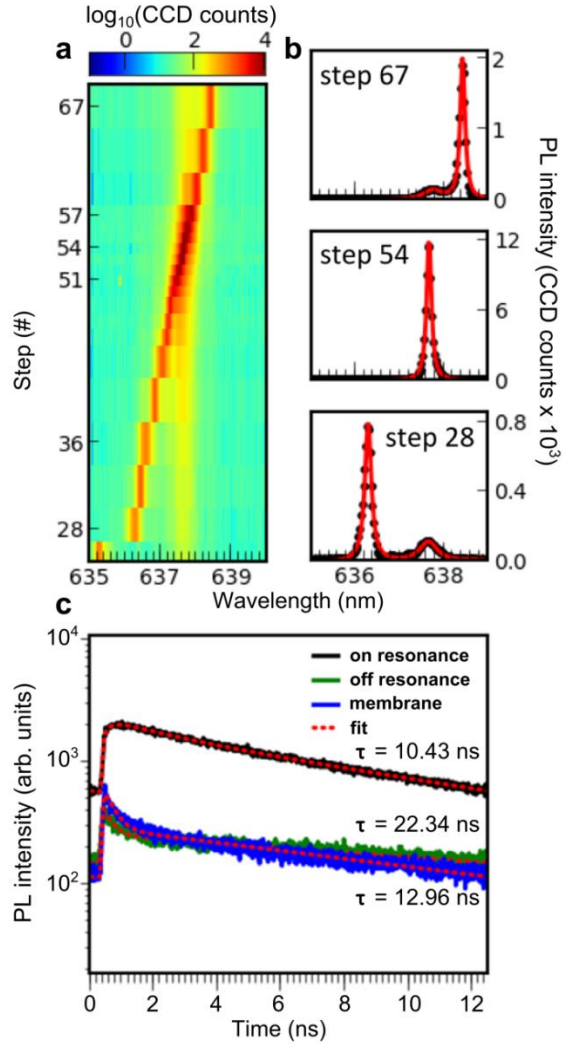

**Figure 3. Resonant enhancement and lifetime reduction of NV zero-phonon line (ZPL). a.** A two-dimensional intensity plot (log scale) of the photoluminescence (PL) spectrum at different cavity tuning steps. **b.** PL spectrum when the cavity mode is off-resonance, on-resonance, and then off-resonance. The cavity mode starts at a wavelength shorter than the NV ZPL (step 28). The intensity of the NV ZPL is enhanced by ~ 40 times when the cavity mode is tuned on resonance (step 54). The NV ZPL intensity then decreases as the cavity mode is tuned off-resonance (step 67). **c.** Lifetime measurements of NV centers are performed when the cavity mode is off resonance (green curve), and on resonance (black curve). NV lifetime is also measured from an un-patterned membrane (blue curve). Lifetime reduction is observed when the NV center is on-resonance with the cavity mode. The lifetime curves are best fit with a double exponential decay (red dotted curve).